\documentstyle [aps,prl,floats,epsf,epsfig]{revtex}

\begin{document}
\draft
\preprint{HEP/123-qed}
\newcommand{\dm}       {\Delta m^2}
\newcommand{\sinq}      {sin^2 2\theta}

\wideabs{
\title{A New Approach to Background Subtraction in Low-Energy Neutrino Experiments}

\author{Y-F.~Wang, L.~Miller, G.~Gratta}
\address{Physics Department, Stanford University, Stanford CA 94305}

\date{\today}
\maketitle
\begin{abstract}
We discuss a new method to extract neutrino signals in low energy
experiments.
In this scheme the symmetric nature of most 
backgrounds allows for direct cancellation from data. 
The application of this technique to the Palo Verde reactor
neutrino oscillation experiment allowed us to reduce the measurement errors
on the anti-neutrino flux from $\sim 20$\% to $\sim 10$\%.    We expect this method 
to substantially improve the data quality in future low background
experiments such as
KamLAND and LENS.

\end{abstract}


}

\narrowtext

\section{introduction}

Backgrounds are a major concern in low-energy neutrino experiments
where signals have low rates and are easily mimicked by other phenomena.
Several types of coincidence schemes, specific to particular neutrino-induced 
processes, have been proposed to improve the signal-to-noise ratio.
One classic example is the use of the 
inverse-$\beta$ decay process
\begin{equation} 
\rm\bar\nu_e + p \longrightarrow e^{+} + n 
\label{eq:inv_beta}
\end{equation}
in liquid scintillator in the discovery of neutrinos and many subsequent 
experiments~\cite{reines,hist}. 
Positrons deposit their energy in the scintillator and annihilate,
yielding two 511 keV $\gamma$'s. Neutrons are captured after thermalization, 
producing $\gamma$'s. The two parts of the event are separated in time by
a delay ranging from tens to hundreds of microseconds depending upon the 
nucleus on which the neutron captures.
The use of similar time correlations has been proposed for 
solar neutrino detection~\cite{LENS}.

It is often the case that experiments are still background limited even
when such coincidence schemes are adopted, particularly 
when the signal rate is very low and can not be varied.
Using data from the Palo Verde neutrino oscillation experiment~\cite{PV_PRL},
we have found 
that most backgrounds have a peculiar symmetry in the energy depositions 
between the two parts of an event that is not present in the neutrino signal.
Such symmetry allows one to eliminate most of the background by direct 
subtraction with the data itself.

In this paper we discuss in detail the method using the data from the Palo Verde 
experiment as an example and 
its application to future experiments such as 
KamLAND~\cite{KamLAND} and LENS~\cite{LENS}, where signal rates
are expected to be substantially lower.

\section{Backgrounds to reactor neutrino experiments}

Low-energy electron anti-neutrinos
from nuclear reactors are unique tools to study
oscillations in the regime of large mixing angle and small mass 
differences. Such combination of parameters has recently received
a good deal of attention as it is consistent with a number of observations
involving solar and atmospheric neutrinos~\cite{solar_atm}. 
In recent times two experiments of this type~\cite{PV_PRL,chooz}
have been set up to search for $\bar\nu_{\rm e}-\bar\nu_{\rm x}$ oscillations 
compatible with the atmospheric neutrino anomaly.
In these experiments, 
electron anti-neutrinos from reactors with energies less than 10 MeV,
are detected by the reaction (1) in
liquid scintillator.
There are two types of backgrounds to this process:
uncorrelated background from environmental radioactivity randomly occurring 
during the time coincidence window, and correlated background from 
cosmic-muon-induced fast neutrons. While the first one can be easily 
measured by varying the time correlation window, the second one is more difficult to 
be measured unambiguously. Neutrons discussed here are produced either in the 
laboratory walls or inside the detector.
Michel electrons from muon decays are not a background since their time 
correlation 
is short and their energy deposition too large.

Fast neutrons can mimic the anti-neutrino signal in the following 
two ways:
\begin{itemize}
\item {\it one-neutron background:} A proton recoil is produced through a 
fast-neutron scattering mimicking the $\rm e^+$ signature; the neutron
is then thermalized and captured like in the case of anti-neutrino events.

\item {\it two-neutron background:} The fast neutron can produce a secondary neutron 
through a spallation process on nuclei; both neutrons are then 
captured simulating the two parts of an anti-neutrino event.

\end{itemize}

Both backgrounds are very difficult to measure except in the case
when the $\bar\nu_{\rm e}$ source (in this case nuclear reactors) 
can be turned off, hence eliminating the signal.   
This
favorable circumstance was available only to the Chooz experiment~\cite{chooz}.
Generally, theoretical models describing neutron production are not considered
accurate enough to provide a viable tool for background subtraction.

Fast neutrons are produced mainly in muon capture and muon spallation. 
While the first process is  well understood, the second is poorly 
known. Although the total neutron yield from muon spallation has been, to some extent,
experimentally measured~\cite{mea,neurate}, theoretical models~\cite{petr,Rya,Allk} are not 
consistent with each other and with data. In addition the few measurements 
of the neutron energy 
spectrum~\cite{karmen} are not well reproduced by
theoretical calculations~\cite{barton,perkins}.
The interpretation of experimental data is complicated by the fact that the neutron
energy spectrum depends upon the muon spectrum that, in turn, is a function of the
depth at which the measurement was carried-on.

Since modern reactor neutrino oscillation experiments typically have long baselines 
and observe anti-neutrinos 
from more than one reactor, in most cases it is impossible to completely 
turn off the signal source. Hence many experiments rely on the power variations 
that generally occur during refueling of some of the reactors in order to subtract
backgrounds.   This method, here referred to as ``ON-OFF method'', has serious 
limitations since: a) the statistical error is large as only a small fraction of the 
neutrinos are used as signal while most of them are subtracted away with the 
background; the smaller the power excursion the larger the statistical error;
b) since reactors are kept to full power for a very large fraction of time (because
of obvious economic reasons) statistical errors are dominated by the short low power
periods while the majority of the data taken by the experiment is not useful to
improve the measurement accuracy; c) the subtraction method only works under
the assumption that backgrounds are stable over the periods of several months 
that separate the full-power periods from the low-power ones; d) complete 
systematic checks on data can only be done after an entire reactor cycle that 
generally corresponds to a period of six months to one year.

The new technique, that we call the ``swap method'', avoids such limitations.

\section{The swap method}

The swap method uses symmetries of the data to directly eliminate 
most of the backgrounds and 
a Monte Carlo calculation to estimate the residual background.
The same symmetries that guarantee the cancellation in data also make the 
whole process rather insensitive to imperfections of the Monte Carlo model.

We first select neutrino events by requiring the prompt part as positron-like
and the delayed part as neutron-like. We have:
\begin{equation}
N_1 = B_{\rm unc} + B_{\rm nn} + B_{\rm pn} + N_{\nu}
\label{eq:a}
\end{equation}
where $N_1$ is the number of selected events,
      $B_{\rm unc}$ the uncorrelated background from natural radioactivity,
      $B_{\rm nn}$ the correlated background from two-neutron captures,
      $B_{\rm pn}$ the correlated background from single-neutron-induced events and 
      $N_{\nu} $ the anti-neutrino signal to be measured.
We then reverse the selection by imposing neutron cuts on the prompt part and
positron cuts on the delayed part, obtaining:
\begin{equation}
N_2 = B'_{\rm unc} + B'_{\rm nn} + \epsilon_1 B_{\rm pn} + \epsilon_2 N_{\nu} 
\label{eq:b}
\end{equation}
Since both uncorrelated and two-neutron backgrounds are symmetric 
under this selection swap, we have $B'_{\rm unc}=B_{\rm unc}$ and $B'_{\rm nn}=B_{\rm nn}$.
Indeed both $B'_{\rm unc}$ and $B_{\rm unc}$ can be measured independently and
are found to be
the same in Palo Verde data.
The terms $B_{\rm pn}$ and $N_{\nu}$ are not {\it a priori} symmetric and we use the factors
$\epsilon_1$ and $\epsilon_2$ to describe the efficiencies for the swapped
selection.

It is essential to realize here that this procedure can only be applied if the 
the trigger system treats the two parts of the event in an identical
fashion. 
At Palo Verde the symmetric 
trigger conditions used~\cite{trigger} were found to have 
an efficiency very similar to the one that would be obtained by separately 
optimizing the
patterns for the positron and the neutron parts of the events.

We can now calculate the difference
\begin{equation}
N_1-N_2 = (1-\epsilon_1) B_{\rm pn} + (1-\epsilon_2) N_{\nu}$$
\label{eq:c}
\end{equation}
where $B_{\rm unc}$ and $B_{\rm nn}$ have been eliminated and $\epsilon_2$ 
can be easily obtained from the $\bar\nu_{\rm e}$ Monte Carlo simulation since 
this process is well known.     The derivation of $\epsilon_1$ is more 
involved since, as already discussed, the neutron background is not 
easy to model.    Here we remark, however, that the ``swap method'' owes its
power to the fact that, as it will be shown later, $\epsilon_1 \sim 1$ and 
$\epsilon_2\sim 0$.
A small $(1-\epsilon_1)$ relaxes the accuracy requirements on the Monte Carlo simulation
to be used to estimate $(1-\epsilon_1)B_{\rm pn}$.

\begin{figure}[tbh!!!!!]
\centerline{\epsfxsize=2.7in \epsfbox{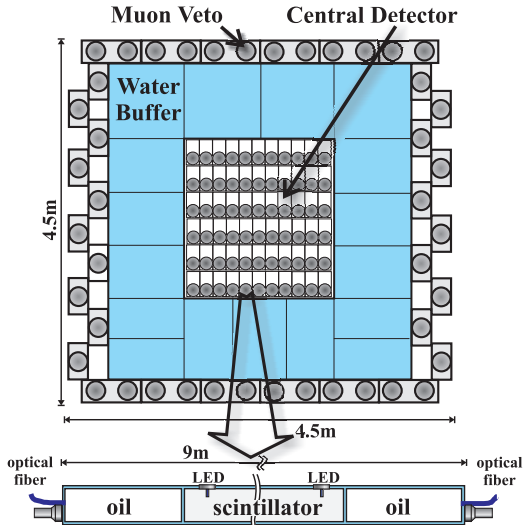}}
\caption{Schematic view of the Palo Verde neutrino detector. The liquid scintillator 
is loaded with 0.1\% Gd in order to reduce the neutron capture time to 30 $\mu$s and 
provide a large capture signal (8 MeV $\gamma$ cascade).}
\label{fig:detector}
\end{figure}

The Palo Verde detector~\cite{prd} is shown in Fig.~\ref{fig:detector}.
All our simulations use the Monte Carlo program GEANT~\cite{geant} to describe the 
detector and the materials surrounding it. 
Electromagnetic interactions 
are simulated by GEANT while hadronic interactions are simulated by 
GFLUKA~\cite{fluka}. Low energy neutron transport is simulated by 
GCALOR~\cite{gcalor}.  Cuts for tracking neutrons are set to 1~MeV 
for concrete and earth, 100~keV for the veto scintillator, 10~keV for the 
water shielding and $10^{-5}$ eV for the central detector.
Light quenching for protons in liquid scintillator is also included~\cite{quench}.  
Our program successfully simulates the behavior of neutrons from Am-Be and 
$\rm e^+$ and $\gamma$'s from $^{22}$Na sources, which proves that 
neutrino signals are simulated correctly~\cite{prd}.

In estimating $B_{\rm pn}$ we consider both the process of muon 
spallation and capture.    
Each process may happen either in the laboratory walls or inside the boundaries
of the veto system. 
Our selection cuts for the positron part of the $\bar\nu_{\rm e}$ event have been
found to have negligible efficiency for muon-induced radioactivity, so that this
phenomenon represents a negligible fraction of the background and is not analyzed
further. Neutrons from other processes such as neutrino interactions with the rock, 
photo-nuclear reactions associated with electromagnetic showers generated by muons, 
and muon elastic scattering, are also found to be negligible.

\subsection{Muon spallation inside laboratory walls}

As it turns out, muon spallation in the concrete walls of the laboratory is the
dominant component of $B_{\rm pn}$ in the case of Palo Verde.
Although absolute rate predictions are not particularly
reliable, in our case a normalization point can be obtained from data where the
prompt part of the event has an energy in excess of 10~MeV.    In fact, above this
energy there is no anti-neutrino signal or neutron capture  
but only proton recoils from neutron collisions.
So we use the simulation only to obtain the ratio
\begin{equation}
r = {B^{\rm MC}_{\rm pn}(E<8~{\rm MeV}) \over B^{\rm MC}_{\rm pn}(E>10~{\rm MeV})}
\label{eq:r}
\end{equation}
and then find  $B_{\rm pn}$ normalizing to data:
\begin{equation}
B_{\rm pn} = r\cdot B^{\rm data}_{\rm pn}(E>10~{\rm MeV}).
\label{eq:B}
\end{equation}
We then determine $(1-\epsilon_1)r$
using the Monte Carlo simulation.

As mentioned above,
the energy spectrum of neutrons from muon spallation 
is not very well known and a broad range of results can be found in the literature.
Barton~\cite{barton} suggested that the spectrum of neutrons from hadronic cascade
follows $E^{-1/2}$ between 10-50 MeV, while the spectrum of neutrons from 
$\pi^-$ capture follows a flat spectrum up to 100 MeV.   
Perkins~\cite{perkins} suggested that the neutron spectrum from muon spallation follows 
$ E^{-1.6}$. The combination of $ (9.7E^{-1/2}+6.0e^{-E/10})~$
has been used in a measurement~\cite{neurate} at a shallow site. 
It has also been suggested~\cite{khalchukov} to use proton and neutron spectra following
$ E^{-1.86}$ as  measured at accelerators 
for photo-nuclear interactions. Finally the Karmen 
experiment reported a visible energy spectrum following $ e^{-E/39}$ 
for spallation neutrons~\cite{karmen}.

We conservatively choose four spectra, including all the options described, 
as input to our Monte Carlo calculation of backgrounds.
Tab.~\ref{tab:spalab} shows
the two extreme cases of $E^{-0.5}$ and $E^{-2.0}$ together with the exponential 
spectrum $e^{-E/39}$ (Models A). We assume that neutrons are produced isotropically.
In addition we compute the neutron spectrum by producing cosmic muons 
in the energy range 0.01~MeV~$<E_{\mu}<$~500~GeV according to the proper energy and 
angular distributions~\cite{garser}.   We then generate real bremsstrahlung $\gamma$'s
according to the distribution $1/E_{\gamma}$ in the energy range 10~MeV~$<E_{\gamma}<E_{\mu}$.
Neutrons are then produced from photo-nuclear processes with a spectral shape 
$E^{-1.86}$~\cite{khalchukov} and an angular distribution from ~\cite{angular}
in the energy range 10~MeV~$<E_{\rm n}<E_{\gamma}$. 
The result of this method is also reported in Tab.~\ref{tab:spalab} as Model B.

\begin{table}[htb]
\begin{center}
\begin{tabular}{|l|c|c|c|c|}
{\rm Model} & $\epsilon_1$       & $r$ & $(1-\epsilon_1)r$   & $B^{\rm MC}_{\rm pn}({\rm d^{-1}})$\\
 & & & & $E>10$~MeV \\
\hline
$E^{-0.5}$, A &  $1.16 \pm 0.07$ & $0.69\pm 0.04$ & $-0.11 \pm 0.05 $& 155 \\
$E^{-2.0}$, A &  $1.20 \pm 0.11$ & $0.67\pm 0.07$ & $-0.13 \pm 0.07 $& 1.7 \\
$e^{-E/39}$, A &  $1.06 \pm 0.07$ & $0.77\pm 0.05$ & $-0.05 \pm 0.06 $& 17 \\
$E^{-1.86}$, B &  $1.15 \pm 0.06$ & $0.76\pm 0.04$ & $-0.11 \pm 0.04 $& 32 \\
\hline
Average  &  &  & $-0.10 \pm 0.05  $ &  \\
\end{tabular}
\vskip 0.2cm
\caption{Results of Monte Carlo simulation for neutrons produced in the laboratory walls
by muon spallation.
The errors shown are due to limited Monte Carlo statistics. The estimated background
rate above 10~MeV 
shown in the last column refers to an independent calculation, described in the text.
The values of $B^{\rm MC}_{\rm pn}(E>10~{\rm MeV})$ should be compared with a total rate of 
$13.5\pm 0.4$~d$^{-1}$ obtained from Palo Verde data. The differences between Models are 
discussed in the text.}
\label{tab:spalab}
\end{center}
\end{table}

In order to obtain the results in Tab.~\ref{tab:spalab}, we use only interactions in a
1~m thick concrete shell as neutrons produced at larger depths are completely absorbed.
The 10~MeV low-energy cutoff used in the calculations is justified by the fact that softer
neutrons are completely absorbed by the 1~m thick water buffer surrounding the Palo Verde
central detector. From Tab.~\ref{tab:spalab} we can see that the proton-recoil 
energy is only weakly dependent upon the neutron energy so that both $\epsilon_1$ and $r$ 
remain almost constant for drastically different neutron energy spectra. 
Furthermore, the uncertainties on $\epsilon_1$ and $r$ 
have little effect on the factor $(1-\epsilon_1)r$ that directly enters the 
neutrino measurement. This implies that the neutron-capture signal, common to both the
neutrino signal and the $B_{\rm pn}$ background, 
is similar to the
proton-recoil signal of the background, but different from the positron
signal of a neutrino event.

In Fig. 2 we show the energy deposited by the neutron-induced 
proton recoil in the most energetic cell of the prompt part of the events. 
The four different neutron spectra used are normalized to data for energies
above 10~MeV.    

\begin{figure*}[htb!!!]
\mbox{\epsfig{file=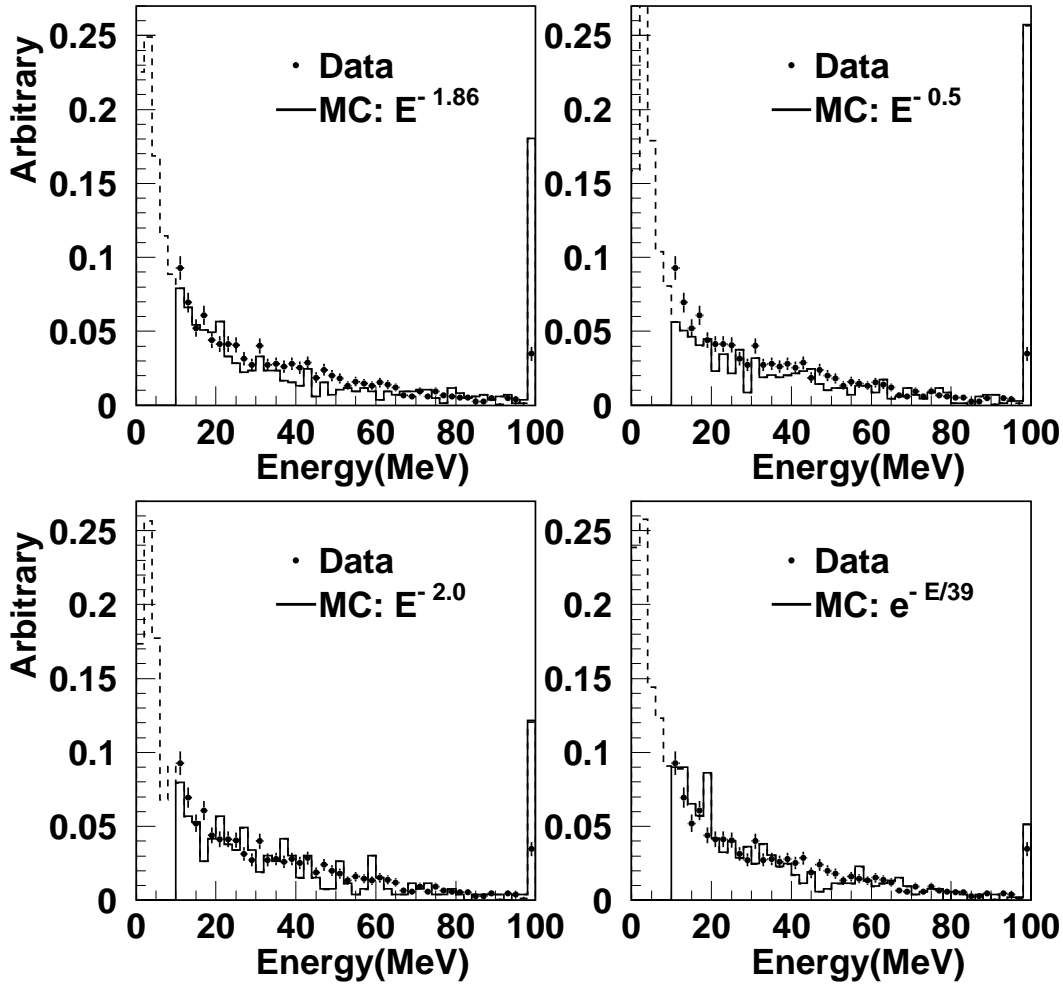,width=17cm,clip=}}
\vskip -1.5 cm
\caption{Energy deposited in the most energetic cell for different neutron spectra. 
Data and Monte Carlo total rates are normalized above 10~MeV where there is
no $\bar\nu_{\rm e}$ signal in the data.  The dashed line is the MC prediction for $E<10$~MeV.}
\label{fig:compare}
\end{figure*}

To verify the results obtained we independently calculate the spallation background by using
a neutron yield of $6\times 10^{-5}~{\rm\mu^{-1} g^{-1} cm^{2}}$ for normalization.
This number is 
obtained by rescaling the measurements of~\cite{neurate} to our depth of 32~m.w.e. 
A total of $7.8\times 10^6$ neutrons are generated
daily in our lab walls. The calculated background rates 
$B^{\rm MC}_{\rm pn}(E>10~{\rm MeV)}$ are shown in the last 
column of Tab.~\ref{tab:spalab},  which can be compared with our measurement
$B^{\rm data}_{\rm pn}(E>10~{\rm MeV}) = 13.5\pm 0.4(stat.)~{\rm d^{-1}}$. It is clear that 
our measurement falls somewhere in the middle of the predictions 
and the spectra chosen for the simulation cover a conservative range of possibilities.
We conservatively maintain all 
four options and use the possible differences as contributions to the systematic errors.

Finally we average $(1-\epsilon_1)r$ from Tab.~\ref{tab:spalab} obtaining $-0.10\pm 0.05$,
and then proceed to calculate the background from neutron spallation in the walls 
as  $(1-\epsilon_1)B_{\rm pn} = 1.35 \pm 0.68~{\rm d}^{-1}$.

\subsection{Muon spallation inside the veto system}

Inefficiencies of the cosmic-ray veto system result in a component of $B_{\rm pn}$
from neutrons produced within the detector. The veto inefficiency at Palo Verde
is measured to be $(0.07\pm 0.02)\%$ 
for through-going muons (two missed hits). 
Only neutrons produced in the water buffer are important here since muons responsible
for neutron spallation directly in the central detector scintillator would be 
easily detected and discarded.

Using the same procedure as above we obtain a neutron yield of
1600~d$^{-1}$ from the water buffer. The corresponding $\epsilon_1$, $r$ and 
$(1-\epsilon_1)r$ are given the in Tab.~\ref{tab:spawater} together with 
background estimates $\mathrm B^{MC}_{pn}(E>10~MeV)$. 
It can be readily 
seen from the Table that  $\mathrm B^{MC}_{pn}(E>10~MeV)$
in the water buffer has a negligible rate compared to that in Tab.~\ref{tab:spalab}, 
and their $(1-\epsilon_1)r$ are very similar.
Hence in the rest of our calculations
we will neglect this contribution.

\begin{table}[t]
\begin{center}
\begin{tabular}{|c|c|c|c|c|c|}
Model & $\epsilon_1$  &   $r$ & $(1-\epsilon_1)r$ &  $B^{\rm MC}_{\rm pn}$ (d$^{-1}$)\\
 & & & & $E>10$~MeV \\
\hline
$E^{-0.5}$, A & $1.17 \pm 0.12$ & $0.72\pm 0.07$ & $-0.12 \pm 0.08  $ & 2.2  \\
$E^{-2.0}$, A & $0.94 \pm 0.08$ & $1.41\pm 0.15$ & $0.08 \pm 0.11  $  & 0.06 \\
$e^{-E/39}$, A & $0.97 \pm 0.04$ & $1.12\pm 0.05$ & $0.03 \pm 0.04  $ & 0.8  \\
$E^{-1.86}$, B & $1.13 \pm 0.05$ & $1.12\pm 0.05$ & $-0.15 \pm 0.06  $& 0.9  \\
\end{tabular}
\vskip 0.2cm
\caption{Results of Monte Carlo simulation for neutrons produced in the 
water buffer by muon spallation.
The errors shown are due to limited Monte Carlo statistics. The estimated rate above 10~MeV 
shown in the last column refers to an independent calculation, described in the text.
It is clear that the rates found for this channel are 
negligible with respect to the rates in Tab.~\ref{tab:spalab}.}
\label{tab:spawater}
\end{center}
\end{table}

\subsection{Muon capture inside laboratory walls}

The muon capture process is rather well understood and the resulting neutrons 
tend to have a soft spectrum (compared to spallation), with an upper limit around
100~MeV (muon mass).
The underground laboratory at Palo Verde is built with low activity concrete using
dolomite as aggregate.  The elemental composition of concrete is shown in Tab.~\ref{tab:cap} 
together with the muon capture rate and the neutron yield per capture 
for each element. 
Almost every capture produces one neutron. 
The total un-vetoed muon rate in the walls is 2~kHz, of which 0.9~kHz is due 
to $\mu^-$. The stopping $\mu^-$ rate amounts to 90 Hz. 
Using Tab.~\ref{tab:cap} we obtain a total muon capture probability
of 67\%, resulting 
a neutron production rate of 60~Hz in the laboratory walls.

\begin{table}[hb!!]
\begin{center}
\begin{tabular}{|c|c|c|c|}
Element & Fraction by &    Capture rate     &    n yield/capture      \\
        &  mass (\%)  & ($10^{5}$~s$^{-1}$) & ($\mu$-capture$^{-1}$) \\
\hline
H  & 0.6  & 0.004~\cite{suzuki}  &   1 \\
C  & 10.4 & 0.388~\cite{suzuki}  &  $\sim 1$~\cite{table} \\
O  & 50.6 & 1.026~\cite{suzuki}  &  0.98~\cite{table} \\
Al & 0.3  & 7.054~\cite{suzuki}  &  1.26~\cite{muk} \\
Si & 1.2 & 8.712~\cite{suzuki}  &  0.86~\cite{muk} \\
Mg & 10.7 & 10.67~\cite{suzuki}  &  1 $^*$\\
Ca & 22.9  & 25.57~\cite{suzuki}  &  0.75~\cite{muk} \\
Fe & 3.3  & 44.11~\cite{suzuki}  &  1.12~\cite{muk} \\
\end{tabular}
$^*${actual value not known, assumed to be 1}
\vskip 0.2cm

\caption{Elemental composition together with muon capture rates and 
neutron yields for the
Palo Verde concrete.   The concrete contains 3\% reinforcing steel, 16\% cement and 81\% crushed 
dolomite aggregate. }
\label{tab:cap}
\end{center}
\end{table}

The neutron energy spectrum from capture is simulated taking into account both the soft 
neutrons from nuclear evaporation and the hard neutrons from direct emission.
For light elements such as $^{12}$C and $^{16}$O, individual lines are present in 
the neutron 
spectrum~\cite{capt,book}, while for heavier elements such spectrum has the 
properties of
a continuum.  We use the energy spectra in ref.~\cite{capt} to simulate neutrons from
$^{12}$C and $^{16}$O, while those from heavy elements are simulated according to~\cite{muk}.
Capture on hydrogen happens at a negligible rate and is disregarded here. 
From the simulation, we obtain $\mathrm B_{pn} = 0.10\pm 0.05~$d$^{-1}$ and
$(1-\epsilon_1)=0.23\pm 0.32$ so that  
$(1-\epsilon_1)B_{\rm pn}=0.02\pm 0.03$~d$^{-1}$ where the error includes Monte Carlo 
statistics and all systematic uncertainties. We conclude that this background 
is negligible compared to other channels.

\subsection{Muon capture inside the veto system}

In analogy to the spallation case, neutrons from muon capture inside the veto 
system can contribute to $B_{\rm pn}$ for un-tagged muons.
Even considering the conclusions from the previous two sections this background
cannot be a-priori dismissed as negligible since the veto counter inefficiency 
for single hits (such as would result from a stopping muon) is measured to be
$(4\pm 1)$\%.   In the buffer-water muon capture on $\rm ^{16}O$ is the only 
significant 
source of neutrons since the capture rate on hydrogen is very small.

The total $\mu^-$ rate in our detector is about 860~Hz, of which 86~Hz are stopping muons. 
This results in a rate of un-tagged neutrons of 52500~d$^{-1}$.
Using the energy spectrum from ref.~\cite{capt}, we obtain from Monte Carlo simulation 
$B_{\rm pn} = 3.9\pm 0.8$~d$^{-1}$ and
$(1-\epsilon_1) = 0.22\pm 0.03$. Finally, this background contributes 
$(1-\epsilon_1)B_{\rm pn}=0.86\pm 0.50$~d$^{-1}$ where, as usual, the error includes all
uncertainties.

\subsection{Verification of the method}

In summary, all the above backgrounds add to a total rate
$(1-\epsilon_1)B_{\rm pn} = 0.5\pm 0.8$~d$^{-1}$, very close to 0. 
The error is dominated 
by systematics, particularly stemming from uncertainties in the neutron energy spectrum. 

In order to verify the correctness of the method, we can directly measure in the data
a similar background by slightly modifying the anti-neutrino selection cuts 
so that no signal is detected.  Positrons (from $\bar\nu_{\rm e}$ interactions) differ
from proton recoils (from background neutrons interactions) by the annihilation $\gamma$'s 
with energies of less than 511~keV.   An event selection requiring more than 600~keV for
each of the hits will result in the total 
rejection of the neutrino signal.   Hence in this case only the $B_{\rm pn}$ term will 
be present after swap selection:
\begin{equation}
N_1-N_2 = (1-\epsilon_1)\cdot B_{\rm pn} 
\label{eq:f}
\end{equation}
Tab.~\ref{tab:con} shows the result of this test with background only.  The values of 
$N_1-N_2$ from data is consistent, within errors, with Monte Carlo estimate of  
$(1-\epsilon_1)B_{\rm pn}$. Different selection cuts for the background 
yield similar results.

\begin{table}[t!]
\begin{tabular}{|l|c|}
                                           & Rate (d$^{-1}$)    \\
\hline
$(1-\epsilon_1)B_{\rm pn}$ (Spallation in walls)       & $0.19\pm 0.26 $    \\
$(1-\epsilon_1)B_{\rm pn}$ (Spallation in detector)    &  -      \\ 
$(1-\epsilon_1)B_{\rm pn}$ (Capture in walls)        &  -      \\ 
$(1-\epsilon_1)B_{\rm pn}$ (Capture in detector)     & $-0.08 \pm 0.08$   \\ 
\hline
Total $(1-\epsilon_1)B_{\rm pn}$ (MC)                &  $0.11\pm 0.27$ \\
\hline
$N_1$ (Data)                                   &  $8.75\pm 0.28$ \\
$N_2$ (Data)                                   &  $9.07\pm 0.29$ \\
$N_1-N_2$ (Data)                                   &  $-0.32 \pm 0.20$ \\
\end{tabular}
\vskip 0.2cm
\caption{Comparison of data and Monte Carlo for an event selection with no efficiency
for the anti-neutrino signal (see text). Errors are statistical only. Note that
$N_1$ and $N_2$ are correlated.}
\label{tab:con}
\end{table}

\section{Comparison with ``ON-OFF'' background subtraction}

The advantages of the swap method become clear in the comparison with
the ``ON-OFF'' method, as shown in Tab.~\ref{tab:resu}.  The Table summarizes
the results of the Palo Verde experiment for the 1999 data taking period from~\cite{PV_PRL}.
$\epsilon_1$ is indeed very close to 1 resulting in a very small residual background
$(1-\epsilon_1)B_{\rm pn}$, even for a background rate $B_{\rm unc}+B_{\rm nn}+B_{\rm pn}$ 
as high as 27~d$^{-1}$.   In the new variable $N_1 - N_2$, not only the terms $B_{\rm unc}$ 
and $B_{\rm nn}$ drop completely, but also $B_{\rm pn}$ is strongly suppressed.
A conservative 160\% uncertainty on $(1-\epsilon_1)B_{\rm pn}$ only corresponds to a 
4\% error on $N_{\nu}$. 
On the other hand, $\epsilon_2$ is only 0.16, small enough that the statistical power 
of the $\bar\nu_{\rm e}$ signal
is retained.

\begin{table}[htb!]
\begin{tabular}{|l|c|c|}
               & 1999 ``ON'' & 1999 ``OFF''\\
\hline
No. of days                &  110.95 &   23.40    \\
$\bar\nu_e$ efficiency     & 0.112   & 0.111      \\
$\mathrm \epsilon_2$       & 0.159 & 0.159    \\
\hline
$(1-\epsilon_1)B_{\rm pn,~spall.~in~walls}$(d$^{-1}$)   & $-1.35\pm 0.68$  & $-1.33\pm 0.67$  \\
$(1-\epsilon_1)B_{\rm pn,~spall.~in~water}$(d$^{-1}$)   & -  & -  \\ 
$(1-\epsilon_1)B_{\rm pn,~capt.~in~walls}$(d$^{-1}$)   & $0.02\pm 0.03 $  & $0.02\pm 0.03 $  \\ 
$(1-\epsilon_1)B_{\rm pn,~capt.~in~water}$(d$^{-1}$)   & $0.86\pm 0.43 $  & $0.86\pm 0.43 $  \\ 
\hline
$\mathrm N_1$ (d$^{-1}$)                                & $52.9\pm 0.7$ & $43.9 \pm 1.4$  \\
$\mathrm N_2$ (d$^{-1}$)                                & $32.3\pm 0.5$ & $31.7 \pm 1.2$  \\
$\mathrm N_{\nu}$ (d$^{-1}$)                            & $25.2\pm 0.9$ & $15.1 \pm 1.9$  \\
\hline
$B_{\rm unc}+B_{\rm nn}+B_{\rm pn}$ (d$^{-1}$)         & $27.7\pm 0.6$ & $28.8 \pm 1.3$  \\
\hline
$\bar\nu_{\rm e}$ observed (d$^{-1}$)                   & $225\pm 8$      & $136   \pm 17$    \\
$\bar\nu_{\rm e}$ expected (d$^{-1}$)               & 218             & 130               \\
\end{tabular}
\vskip 0.2cm
\caption{Palo Verde results from 1999 data taking. Errors are statistical except 
for $(1-\epsilon_1)B_{\rm pn}$ where errors are systematic. The individual background rates are 
approximately 4~d$^{-1}$ for $B_{\rm unc}$, 14~d$^{-1}$ for $B_{\rm nn}$ and 10~d$^{-1}$ for 
$B_{\rm pn}$}
\label{tab:resu}
\end{table}

Correcting $\mathrm N_1$ in both columns by their respective efficiencies and  
subtracting column 2 from column 1 in the Table, we find that the ``ON-OFF'' method gives 
a neutrino rate of $77\pm 14({\rm stat.}) \pm 8({\rm syst.})$~d$^{-1}$ for an expectation
of 88~d$^{-1}$ in the no-oscillation hypothesis.    In calculating the signal essentially
only one reactor out of three is used, while the statistical fluctuations in the flux of
all reactors along with the background contribute to the errors.  The systematic error 
includes uncertainties on positron and neutron efficiencies (5\%), $\bar\nu_e$ 
selection (8\%) and $\bar\nu_e$ flux estimate (3\%).
 
In the case of the swap method, we find
$225\pm 8 ({\rm stat.}) \pm 17({\rm sys.})$~d$^{-1}$ 
($137\pm 17({\rm stat.}) \pm 14({\rm sys.})$~d$^{-1}$) for high (low) power against a 
prediction of 218~d$^{-1}$(130~d$^{-1}$) for the case of no oscillations. Here all reactors
contribute to the signal and, in fact, the contributions for the two periods with different
power can be used together to strengthen the measurement.   Indeed the statistical error
drops from 18\% in the case of ``ON-OFF'' to 3.5\% (12\%) for high (low) power.   
While systematic errors from efficiencies and flux are the same as in the previous case, 
the error on $\rm\bar\nu_e$ selection is now only 4\% because some of the selection systematics 
cancel in the $N_1-N_2$ difference.  A new uncertainty due to the $B_{\rm pn}$ estimate 
(4\%) appears.

\section {Summary and discussions}

We have shown that a novel method of analysis, applicable to low energy neutrino
experiments using correlated signatures, provides substantially more accurate 
background subtraction over more traditional techniques. 
While the new method was applied first to a reactor neutrino oscillation experiment,
it can be more generally used in experiments where: a) the signal events consist of 
two sub-events correlated in time or space, b) the two sub-events are distinctively 
different from each other in signal but similar in backgrounds or vice-versa, and 
c) the detector and trigger treat the two sub-events in identical fashion.

These criteria apply to several future neutrino experiments such as KamLAND and LENS.   
In KamLAND~\cite{KamLAND} electron anti-neutrinos from reactors will be 
detected in 1~kton liquid scintillator as a positron with energy deposit between 1 and 
8~MeV correlated in time with a neutron which gives a 2.2~MeV $\gamma$ line from capture
on protons. The correlated neutron background, which includes both 
one- and two-neutron events, is expected to have the same magnitude as the random 
background. The application of the method is therefore 
straightforward: both random and two-neutron backgrounds can be eliminated 
and the one-neutron background can be estimated in a way which is very similar to
 what we discussed above.
The LENS experiment~\cite{LENS} is designed to detect solar neutrinos via, for example,
the process 
$\nu + ^{160}{\rm Gd} \rightarrow {\rm e^-} + ^{160}{\rm Tb^*}$, where the signature 
consists of an 
electron (0.04-2 MeV) and a $\gamma$ (64 keV) correlated in time.
One of the main backgrounds is the random coincidence of $\gamma$'s from radioactive
impurities in the detector and it can be easily suppressed by the method described above.

\section {Acknowledgments}

We would like to thank our Palo Verde collaborators for their continuing support and 
constructive criticism.    Special thanks go to J.~Busenitz for cross checks
of the method's dependence on muon spallation models
and P.~Vogel for countless suggestions and advise.   This work was supported
in part by DoE grant DE-FG03-96ER40986.   One of us (L.M.) would like to thank the ARCS
foundation for its generous support.

\end{document}